\documentstyle[multicol,aps,prl]{revtex} 

\renewcommand{\narrowtext}{\begin{multicols}{2} \global\columnwidth20.5pc}
\renewcommand{\widetext}{\end{multicols} \global\columnwidth42.5pc}
\def\top#1{\vskip #1\begin{picture}(290,80)(80,500)\thinlines \put(
65,500){\line( 1, 0){255}}\put(320,500){\line( 0, 1){
5}}\end{picture}}
\def\bottom#1{\vskip #1\begin{picture}(290,80)(80,500)\thinlines \put(
330,500){\line( 1, 0){255}}\put(330,500){\line( 0, -1){
5}}\end{picture}}

\def\al{\alpha}
\def\be{\beta}
\def\ga{\gamma}
\def\de{\delta}

\def\et{\eta}

\def\la{\lambda}

\def\ps{\psi}

\def\De{\Delta}

\def\La{\Lambda}
\def\Si{\Sigma}

\def\Ph{\Phi}
\def\Ps{\Psi}

\def\cN{{\cal N}}
\def\cS{{\cal S}}
\def\cO{{\cal O}}
\def\ap{\alpha^\prime}
\def\fr#1#2{{{#1} \over {#2}}}
\def\frac#1#2{\textstyle{{{#1} \over {#2}}}}
\def\pt#1{\phantom{#1}}

\def\ket#1{|{#1}\rangle}
\def\bra#1{\langle{#1}|}

\def\half{{\textstyle{1\over 2}}}
\def\lsim{\mathrel{\rlap{\lower4pt\hbox{\hskip1pt$\sim$}}
    \raise1pt\hbox{$<$}}}
\def\gsim{\mathrel{\rlap{\lower4pt\hbox{\hskip1pt$\sim$}}
    \raise1pt\hbox{$>$}}}

\def\da{\dagger}
\def\ad{{a^\dagger}}
\def\bd{{b^\dagger}}
\def\cd{{c^\dagger}}
\def\sd{{s^\dagger}}
\def\td{{t^\dagger}}
\def\ud{{u^\dagger}}
\def\ti{\tilde}
\def\third{\textstyle{1\over3}}
\def\quarter{\textstyle{1\over4}}
\def\lvac{\langle0|}
\def\rvac{|0\rangle}

\newcommand{\beq}{\begin{equation}}
\newcommand{\eeq}{\end{equation}}
\newcommand{\bea}{\begin{eqnarray}}
\newcommand{\eea}{\end{eqnarray}}
\newcommand{\rf}[1]{(\ref{#1})}

\begin{document}

\title{ Analytical construction of a nonperturbative vacuum
for the open bosonic string }    
\author{V.\ Alan Kosteleck\'y$^a$ and Robertus Potting$^b$}
\address{$^a$Physics Department, Indiana University, 
          Bloomington, IN 47405, U.S.A.}
\address{$^b$U.C.E.H., Campus de Gambelas, Universidade do Algarve, 
          8000 Faro, Portugal}
\date{IUHET 426, August 2000} 
\maketitle

\begin{abstract}

Using analytical methods,
a nonpertubative vacuum is constructed recursively 
in the field theory for the open bosonic string.
Evidence suggests it corresponds to the Lorentz-invariant endpoint
of tachyon condensation on a D25-brane.
The corresponding string field is a twisted squeezed state.

\end{abstract}


\narrowtext

\section{Introduction}
\label{intro}

The physical behavior of strings
can be investigated in a variety of ways.
Much of the existing lore is based on perturbative studies 
of single strings in relativistic quantum mechanics. 
However, 
it is likely that a satisfactory understanding
of the subject requires also mastering 
nonperturbative and collective string behaviors.

String field theory offers one framework within which
to investigate nonperturbative many-body string phenomena.
An essential feature of any field theory 
is the structure of its vacuum, 
and it is of particular interest to establish 
the existence of any nonperturbative vacua.
A relatively simple string field theory 
describes the open bosonic string
\cite{ew},
and for this case the existence of at least one nonperturbative vacuum
has been established
\cite{ksobs,kspc,kp2}.
The procedure involves a level-truncation scheme
in which successive approximations to the full theory are made 
according to the mass level of the particle fields 
and the terms in the action.
Analytical methods at low truncation orders and
a combination of analytical and numerical methods at higher orders 
can determine the structure of the string field $\cN$
for the nonperturbative vacuum
in terms of expectation values of particle fields.

The nonperturbative vacuum 
$\cN$
has been conjectured by Sen to be
the end product of tachyon condensation on a D25-brane
\cite{as}.
Strong support in favor of this conjecture exists 
from explicit calculations using the level-truncation scheme
in the field theory for the open bosonic string
\cite{sz1,wt1,mt},
with these now having reached level (10,20)
in the notation of Ref.\ \cite{kp2}.
Analogous calculations for the superstring
provide further support for the idea that
the tachyon effective potential has
a minimum where the D-brane tension is exactly cancelled
\cite{nb,bsz,dr,jd}.

In this work,
we present an analytical approach to constructing
the nonperturbative vacuum $\cN$.
We obtain a series representation for $\cN$
and compare it to the numerical solution
obtained via the truncation scheme.
The methodology and results offer several interesting possibilities 
for future exploration.
In the remainder of this introduction,
we provide a brief summary of the steps involved
in the construction of $\cN$,
and we offer some motivation for our procedure.

The steps involved in the construction of $\cN$ are as follows.
First,
note that $\cN$ is to be determined
as a set of constant expectation values 
satisfying the string-field equations of motion:
\beq
Q\cN + g \ap \cN \star \cN = 0 .
\label{eqmot}
\eeq
The requirement of spacetime independence of $\cN$ makes
it useful to transform the cubic vertex to
a form in which the oscillator zero modes are converted
into the momentum representation,
so that the momentum may be set to vanish. 
Once this form of the cubic vertex is obtained,
we search for a string field $\cS$ obeying 
\beq
\cS\star \cS = c_0\cS ,
\label{cSdef}
\eeq
which serves as the basic object upon which the solution
is to be constructed.
We next perform a Bogoliubov transformation 
to a new oscillator basis 
for which $\cS$ is the vacuum state.
In this new basis,
the string field $\cN$ can be expanded in particle-field modes
and the conditions determining their constant values can be extracted
from the equations of motion \rf{eqmot}.
These conditions form a recursive set 
that is amenable to a formal solution for $\cN$.
Certain features of the solution can be compared to those
already established via the level-truncation scheme,
with good agreement.

The above procedure may appear somewhat convoluted at first sight,
so we provide some heuristic physical and mathematical motivation 
before detailing the calculations in the following sections.
On the physical side,
one intuitively expects $\cN$ to describe a situation 
in which the whole string field has condensed to the vacuum.
It is therefore plausible that 
an infinite number of particle fields acquire vacuum values
in the nonperturbative vacuum $\cN$.
In fact,
it has been shown  
\cite{aku}
that no nontrivial finite linear combination
of expectation values for the particle fields
can satisfy Eq.\ \rf{eqmot}.
Intuition suggests that the lighter modes should 
play a greater role in the development of the string-field condensate,
so one might anticipate expectation values in $\cN$
to drop with mass level.
This is supported by evidence from
numerical calculations with the level-truncation scheme.
Furthermore,
physical intuition about harmonic oscillators 
and the formation of a coherent condensate in the perturbative vacuum 
also suggests the natural form of the string field $\cN$ is
likely to be closely related to a generalized coherent or squeezed state
\cite{ap}.
These notions are echoed in our construction,
since the solution we obtain for $\cS$ in fact is a squeezed state
and the nonperturbative vacuum $\cN$
is developed by twisting $\cS$ 
with oscillator factors controlled by the BRST operator $Q$.

On the mathematical side,
we observe that the usual description 
in terms of perturbative oscillator modes
and the level-truncation scheme itself
use a basis for the Fock space
in which $Q$ is essentially diagonal
and the star product has a complicated realization. 
However,
the structure of the string field equations
can be viewed as analogous to a Riccati differential equation,
involving a single derivative operator (basically, $L_0 \Ps$),
a linear term (the factor $-\Ps$),
and a quadratic term (the product $\Ps \star \Ps$)
\cite{fn1}.
In such equations
the nonlinear term represents the greatest complication,
and it is therefore natural 
to seek a representation in which the
star product has a relatively simple structure.
The ideal case would be to convert the quadratic term to linear form.
This can be approximated by finding a string state $\cS$
satisfying Eq.\ \rf{cSdef}
and converting to a basis in which $\cS$ is the basic (ground) state,
so that the star product has a relatively simple realization.
The expression for $Q \Ps$ becomes complicated
in the new basis,
so it might seem that little has been gained.
However,
the net effect of the manipulations 
is that the difficulty has been moved from  
the nonlinear part of the equation to the linear part,
which provides just enough advantage to make possible
the construction of a solution.

The remainder of the paper is organized as follows.
In Sec.\ II,
some preliminaries are discussed,
including the conversion of the oscillator zero modes
to the momentum representation 
and some properties of coherent and squeezed states. 
The reader uninterested in these details
may wish to pass directly to section III,
in which the squeezed string field $\cS$ is obtained.
Section IV converts to the squeezed-oscillator basis.
The construction of the nonperturbative vacuum $\cN$ is
presented in Sec.\ V,
along with some of its properties.
The results are discussed in Sec.\ VI.
Finally,
Appendix A contains a derivation of a useful identity. 
Throughout much of this work,
the string coupling $g$ as defined in Ref.\ \cite{ksobs}
and the string tension $\ap$ are set to one,
although they are explicitly displayed in 
certain formulae for clarity.

\section{Preliminaries}

\subsection{Momentum representation}

The equations of motion of the string field 
can be expressed in terms of particle fields
using a Fock-space representation
\cite{gj,cst,ss}.
We are interested in spacetime-independent solutions of these equations,
so it is useful to express the spacetime dependence 
of the vertex in the momentum representation 
rather than in the representation with 
oscillator zero modes.
Following the approach of Ref.\ \cite{gj},
we require the three-vertex $V_3$ 
involving the vertex functions $V'{}^{rs}_{mn}$
to satisfy 
\widetext
\top{-2.8cm}
\hglue -1 cm 
\bea
V_3 &=&
\exp(
-\half \sum_{r,s} \sum_{m,n\ne 0}
a_m^{r\da} V'{}^{rs}_{mn} a_n^{s\da}
-\sum_{r,s} \sum_{n\ne 0} 
a_0^{r\da} V'{}^{rs}_{0n} a_n^{s\da}
-\half \sum_r 
a_0^{r\da} V'{}^{rr}_{00} a_0^{r\da}
)|0\rangle_{123} 
\nonumber\\
&\equiv&
\int dp^1\,dp^2\,dp^3
\exp(
-\half \sum_{r,s} \sum_{m,n\ne 0}
a_m^{r\da} V{}^{rs}_{mn} a_n^{s\da}
-\sum_{r,s} \sum_{n\ne 0} 
p^rV{}^{rs}_{0n} a_n^{s\da}
-\half \sum_r 
p^rV{}^{rr}_{00}p^r
)|0,p\rangle_{123} ,
\eea
\bottom{-2.7cm}
\narrowtext
\noindent  
where the vertex functions
$V^{rs}_{mn}$
determine $V_3$ in the momentum representation of interest. 
As usual,
the $r$, $s$ superscripts are understood to take values modulo 3.  

Rewriting the right-hand side using 
\beq
|p\rangle=(2\pi)^{-1/4}\exp[-\frac 14 p^2+a^\da_0 p
-{\half}(a^\da_0)^2]|0\rangle ,
\eeq
we find the relations
\bea
V'{}^{rs}_{mn}&=&V^{rs}_{mn}
-2\sum_t{V_{m0}^{rt} V_{0n}^{ts}\over 2V_{00}^{tt}+1} , \quad m,n\ne 0 ,
\nonumber\\
V'{}^{rs}_{0n}&=&{2V^{rs}_{0n}\over2V_{00}^{rr}+1} , \quad n\ne 0 ,
\nonumber\\
V'{}_{00}^{rr}&=&{2V_{00}^{rr}-1\over2V_{00}^{rr}+1} .
\eea
Next,
we use these relations to derive some properties of the transformed
vertex functions.

First,
we note that 
the `double square' of the vertex functions 
$V'{}^{rs}_{mn}$ generates the identity
\cite{gj}:
\beq
\sum_{k,t}
V'{}^{rt}_{mk}V'{}^{ts}_{kn}=\delta^{rs}\delta_{mn} , \quad m,n=0,1,\ldots 
\label{Vsquare}
\eeq
This leads to the following identities 
for the vertex functions in the momentum representation: 
\bea
\sum_{k\ge1}\sum_t
V^{rt}_{mk}V^{ts}_{kn}&=&\delta^{rs}\delta_{mn} , \quad m,n=1,2,\ldots ,
\nonumber\\
\sum_{k\ge1}\sum_t V^{rt}_{mk}V^{ts}_{k0}&=&V^{rs}_{m0} , \quad n=1,2,\ldots ,
\nonumber\\
\sum_{k\ge1}\sum_t V^{rt}_{0k}V^{ts}_{k0}&=&2V^{rr}_{00}\delta^{rs}.
\eea
In particular, we see from the first of these relations
that the vertex function $V^{rs}_{mn}$,
for which $m,n$ are restricted to nonzero level numbers,
double-squares to the identity in analogy with Eq.\ \rf{Vsquare}.

Next, 
we establish some features of the structure 
of the transformed vertex functions $V^{rs}_{mn}$.
Recall that 
the vertex functions $V'{}^{rs}_{mn}$ viewed as matrices
can be expressed as
\cite{gj}
\bea
V'{}^{rr}&=&\frac 13 (C+U'+\bar U') ,
\nonumber \\ 
V'{}^{r\,r+1}&=&\frac 13 (C+\alpha U'+\alpha^*\bar U') ,
\nonumber \\ 
V'{}^{r\,r+2}&=&\frac 13 (C+\alpha^* U'+\alpha\bar U') ,
\eea
where $\al=\exp(2\pi i/3)$, 
$C_{mn}\equiv (-1)^m\delta_{mn}$,
and the matrices 
$U'$ and $\bar U' \equiv C\,U'\,C$ satisfy
\beq
(U')^2=(\bar U')^2=1, \quad (U')^\da = U'. 
\label{cupprop}
\eeq
Some algebra shows that these relations imply
\bea
V^{rr}&=&\frac 13 (C+U+\bar U) ,
\nonumber\\
V^{r\,r+1}&=&\frac 13 (C+\alpha U+\alpha^*\bar U) ,
\nonumber\\
V^{r\,r+2}&=&\frac 13 (C+\alpha^* U+\alpha\bar U) ,
\label{expressV}
\eea
where
\beq
U_{mn}=U'{}_{mn}
+(V^{rr}_{00}+ \half) U'{}_{m0}U'{}_{0n} ~,
\quad m,n=1,2,...  ,
\eeq
and $\bar U=C\,U\,C$.
It also follows that 
\beq
U^2=\bar U^2=1, \quad U^\da = U,
\label{cuprop}
\eeq
as before.
These equations imply that
the matrix $V^{rr}\,C = C\,V^{rr}=\frac 13 (1+UC+CU)$ commutes with
all the vertex-function matrices $V^{rs}$, 
while $C\,V^{rs}=V^{sr}\,C$.
These identities are used below.

The above arguments establish that
the vertex functions $V^{rs}_{mn}$ restricted to $n,m\ge1$
have the same formal structure as the original vertex functions
$V'{}^{rs}_{mn}$.
This means that searches for spacetime-independent solutions
to the equations of motion
involve the same formal vertex structure
as spacetime-dependent ones.
We return to this point in section \ref{Discussion}.

In the remainder of this paper
except where otherwise stated,
the indices $m$, $n$ range over $1, 2, \ldots$,
so the results are independent of momentum.
Thus, for example,
the bosonic part of the three-vertex $V_3$ takes the form:
\beq
|V_3^b\rangle=
\exp( -\half\sum_{r,s=1}^3 a^{r\da}_m V^{rs}_{mn}  a^{s\da}_n)\rvac_{123} .
\label{bosonicvertex}
\eeq

\subsection{Identities for coherent and squeezed states}

This subsection presents some identities required
in subsequent sections of the paper,
where frequent use is made of generalized squeezed states.
First, 
recall a basic identity for coherent states in one oscillator dimension,
\beq
\lvac \exp(\la a) \exp(\mu a^\dagger) \rvac = \exp(\la\mu) ,
\label{coherent1d}
\eeq
and the corresponding identity for squeezed states,
\beq
\lvac \exp(\half a\,S\,a) \exp(\half a^\dagger\,V\,a^\dagger) \rvac
= (1-S\,V)^{-1/2} .
\label{squeezed1d}
\eeq
Combining Eqs.\ \rf{coherent1d} and \rf{squeezed1d}
to an identity for displaced squeezed states in one oscillator dimension
yields
\widetext
\top{-2.8cm}
\hglue -1 cm 
\bea
&&\lvac \exp(\la a + \half a\,S\,a)
\exp(\mu a^\dagger + \half a^\dagger\,V\,a^\dagger)  \rvac 
\nonumber\\
&&
\qquad\qquad
\qquad\qquad
\qquad\qquad
= (1-S\,V)^{-1/2}
\exp[\la(1-V\,S)^{-1}\mu + \half\la(1-V\,S)^{-1}V\,\la
+\half\mu(1-S\,V)^{-1}S\,\mu] .
\label{displaced1d}
\eea

The analogous identity for  
$\lvac \exp(\la_n a_n + \half a_m\,S_{mn}\,a_n)
\exp(\mu_n a^\dagger_n + \half a^\dagger_m\,V_{mn}\,a^\dagger_n) 
\rvac$ 
in the multidimensional case is expected to have the form
\bea
&&\lvac \exp(\la\cdot a + \half a\cdot S\cdot a)
\exp(\mu\cdot a^\dagger + \half a^\dagger\cdot V\cdot a^\dagger)  \rvac 
\nonumber\\
&&
\quad\qquad
= 
\hbox{Det}(1-S\cdot V)^{-1/2}
\exp[\la\cdot(1-V\cdot S)^{-1}\cdot\mu + \half\la\cdot(1-V\cdot S)^{-1}
\cdot V\cdot \la+\half\mu\cdot(1-S\cdot V)^{-1}\cdot S\cdot\mu] ,
\label{displacedbosonic}
\eea
\bottom{-2.7cm}
\narrowtext
\noindent  
where the dot indicates contraction of indices.
This can be shown explicitly in two dimensions, 
for example,
by making repeated use of Eq.\ \rf{displaced1d}.

In the fermionic case, 
with an anticommuting oscillator algebra
\cite{fn2}
\beq
\{b_m,c_n^\da\}=\{c_m,b_n^\da\}=\delta_{mn} ,
\quad m,n = 1,2 \ldots ,
\eeq
and all other anticommutators zero,
the equivalent of Eq.\ \rf{displacedbosonic} 
can be found as
\bea
&&
\lvac \exp( -b\cdot \ti S\cdot c)
\exp( b^\dagger\cdot\mu^b + \mu^c\cdot c^\dagger + 
b^\dagger\cdot\ti V\cdot c^\dagger)  \rvac
\nonumber\\
&& ~~= \hbox{Det}(1-\ti S\cdot\ti V)
\exp[\mu^c\cdot(1-\ti S\cdot \ti V)^{-1}\cdot \ti S\cdot\mu^b] .
\label{displacedghost}
\eea
For simplicity, 
we have taken the analogue of the parameter $\lambda$ 
to vanish here.

\widetext
Consider next the star product 
$\cS_1 \star \cS_2$
of two squeezed states,
$\cS_1 \equiv\exp( \half a^{1\da}\cdot S^{11}\cdot a^{1\da})\rvac_1$ 
with squeeze matrix $S^{11}$ and 
$\cS_2 \equiv \exp( \half a^{2\da}\cdot S^{22}\cdot a^{2\da})\rvac_2$
with squeeze matrix $S^{22}$.
The result can be written as
\beq
\ket{\cS_1 \star \cS_2}_3 \equiv
{}_{12}\lvac 
\exp( \half a^1\cdot S^{11}\cdot a^1+\half a^2\cdot S^{22}\cdot a^2)
\exp( \mu^1\cdot a^{1\da} +\mu^2\cdot a^{2\da}+
\half \sum_{r,s=1}^2 a^{r\da}\cdot V^{rs}\cdot a^{s\da})
\rvac_{123} ,
\eeq
where $\mu^1\equiv a^{3\da} V^{31}$, $\mu^2\equiv a^{3\da} V^{32}$.
Evaluating first the expectation value of the 1-oscillators using
\rf{displacedbosonic} on the level-number indices $m, n$ 
and then evaluating the expectation value of the 2-oscillators 
produces a somewhat cumbersome expression.
However,
it simplifies in the special case for which
the matrices $S^{11} C$ and $S^{22}C$ commute with $V^{rs}$.
In the next section,
it is shown that this case is the relevant one
for the construction of solutions of the string field equations.
With the simplification,
the result of the calculation is 
\beq
\ket{\cS_1 \star \cS_2}_3 =  
= [\hbox{Det}_{mn}(\hbox{Det}_{rs}(1-\Si V))]^{-1/2}
\exp[\sum_{r,s = 1}^2\mu^r[(1-\Si V)^{-1}\Si]^{rs}\mu^s]\rvac_3 ,
\label{starsqueezedbosonic}
\eeq
\bottom{-2.7cm}
\narrowtext
\noindent  
where we have introduced a matrix $\Si^{rs}$
in the two-dimensional string indices 
given by  
\beq
\Si^{rs}=\pmatrix{S^{11}&0\cr 0&S^{22}} .
\label{Sigma}
\eeq
In Eq.\ \rf{starsqueezedbosonic},
note the double determinant that occurs in terms of the 
string indices $r,s$ 
and the level-number indices $m,n$.

In the ghost sector,
the three-vertex takes the form 
\cite{fn3}
\beq
|V_3^{gh}\rangle=
\exp[ \sum_{r,s=1}^3 b^{r\da}_m
(E\ti V^{rs} E^{-1})_{mn} c^{s\da}_n]
\ket{+}_{123} , 
\eeq
where the matrix $E$ is given by $E_{mn}= \de_{mn}/\sqrt{m}$.
In this equation,
the matrices $\ti V^{rs}$ have the form
\bea
\ti V^{rr}&=&\third(C+\ti U+\bar{\ti U}),
\nonumber\\
\ti V^{rr+1}&=&\third(C+\al \ti U+\al^*\bar{\ti U}),
\nonumber\\
\ti V^{rr+2}&=&\third(C+\al^*\ti U+\al\bar{\ti U}) ,
\eea
with 
$\bar{\ti U}=C\,\ti U\,C$,
$\ti U^2=\bar{\ti U}{}^2=1$,
and $\ti U^\dagger = \ti U$,
in complete analogy with the bosonic sector. 
Introducing matrices 
$\ti S^{11}$ and $\ti S^{22}$ 
such that $\ti S^{11}C$ and $\ti S^{22}C$ 
commute with $\ti V^{rs}$,
the star product of
two squeezed states is given 
in this case by
\widetext
\top{-2.8cm}
\hglue -1 cm 
\bea
&&{}_{12}\bra{-} \exp(b^1\cdot E \ti S^{11} E^{-1}\cdot c^1
+b^2\cdot E \ti S^{22}E^{-1}\cdot c^2)
\nonumber\\
&&
\qquad \qquad
\times 
\exp( \mu^{c1}E^{-1}\cdot c^{1\da}+ b^{1\da}\cdot E\mu^{b1}+
\mu^{c2}E^{-1}\cdot c^{2\da}+ b^{2\da} \cdot E\mu^{b2}+
\sum_{r,s=1}^2 b^{r\da}\cdot E\ti V^{rs}E^{-1}\cdot c^{s\da})
\ket{+}_{123}
\nonumber\\
&&
\qquad \qquad
\qquad \qquad
\qquad \qquad
=[\hbox{Det}_{mn}(\hbox{Det}_{rs}(1-\ti \Si \ti V))]
\exp[\sum_{r,s=1}^2 \mu^{cr}[(1-\ti \Si \ti V)^{-1}\ti \Si ]^{rs}\mu^{bs}]
\ket{+}_3 .
\label{ssproduct}
\eea
\bottom{-2.7cm}
\narrowtext
\noindent  
In this equation,
we have set 
\beq
\ti \Si^{rs}=\pmatrix{\ti S^{11}&0\cr 0&\ti S^{22}},
\eeq
in analogy with the bosonic sector.

\section{The squeezed string field $\cS$}

In this section,
we obtain squeezed-state solutions of Eq.\ \rf{cSdef}.
The nonperturbative vacuum $\cN$ is constructed
in section V using these solutions.

We have shown in Eqs.\ \rf{starsqueezedbosonic} and \rf{ssproduct} 
that the star product of two squeezed states
is again a squeezed state.
To solve Eq.\ \rf{cSdef},
we require invariance of the matrix defining the
width of the squeezed state.
In the bosonic sector,
this leads to the condition
\beq
C S C= \pmatrix{V^{12},&V^{21}}(1-\Si V)^{-1} 
\Si \pmatrix{V^{21}\cr V^{12}} +V^{11},
\label{eqS11}
\eeq
where we have chosen  
\beq
\Si^{rs}=\pmatrix{S&0\cr 0&S}.
\eeq
In Eq.\ \rf{eqS11},
the matrices $V^{12}=V^{31}$ and $V^{21}=V^{32}$
and their conjugates
arise from the coefficients $\mu_r\equiv a^{3\da} V^{3r}$ in
\rf{starsqueezedbosonic}.
The term $(1-\Si V)^{-1}\Si$ is to be interpreted
as carrying both level-number indices $m$, $n=1,2,\ldots$ 
and string indices $r,s=1,2$.
The $C$ matrices multiplying $S$ on the left-hand side
emerge from the application of the two-vertex on the original state.

The challenge is to solve Eq.\ \rf{eqS11} for $S$,
thereby determining the width of the squeezed state
satisfying Eq.\ \rf{cSdef}.
It is convenient first to develop some machinery
controlling the commutation of the various matrices 
on the right-hand side of Eq.\ \rf{eqS11}.

Using the decomposition of the vertex functions $V^{rs}$ in terms
of the matrices $C$ and $U$, 
the following identities can be shown to hold: 
\bea
V^{rs}C&=&C V^{sr},
\nonumber \\
\pmatrix{V^{12},&V^{21}}\pmatrix{V^{11}&V^{12}\cr V^{21}&V^{11}}&=&
-V^{11}\pmatrix{V^{12},&V^{21}},
\nonumber \\
\pmatrix{V^{21},&V^{12}}\pmatrix{V^{11}&V^{12}\cr V^{21}&V^{11}}&=&
2V^{11}\pmatrix{V^{12},&V^{21}}
\nonumber \\
&&+C\pmatrix{V^{21},&V^{12}} .
\label{relations}
\eea
To simplify the structure of the calculation,
we introduce a two-dimensional space 
\beq
\pmatrix{A\cr B}
\equiv
A\pmatrix{V^{12},&V^{21}}+B\pmatrix{V^{21},&V^{12}} ,
\label{2dspace}
\eeq
where $A$ and $B$ are scalars 
assumed to commute with all other expressions.
In terms of this formalism,
the relations \rf{relations} are 
\bea
\pmatrix{A\cr B}C&=&C\pmatrix{B\cr A},\label{ABc}\\
\pmatrix{A\cr B}V&=&\pmatrix{-V^{11}&2V^{11}\cr 0&C}\pmatrix{A\cr B}.
\label{ABV}
\eea
These equations can be used to determine the
result of commuting the $C$ and $V$ matrices 
in Eq.\ \rf{eqS11} to the left.

As a further simplification,
we make the ansatz 
\beq
S = C T ,
\label{ansa}
\eeq
where $T$ is taken to commute
with $V^{rs}$.
This implies only the product $CV$ need be moved to the left
in Eq.\ \rf{eqS11}.
Combining Eqs.\ \rf{ABc} and \rf{ABV} gives
\beq
\pmatrix{A\cr B}CV=\pmatrix{2V^{11}C&-V^{11}C\cr 1&0}\pmatrix{A\cr B}
\equiv \La\pmatrix{A\cr B}.
\eeq
Commuting $(1-\Si V)^{-1}=(1-CTV)^{-1}$ to the left yields $(1-T\La)^{-1}$,
which can be explicitly evaluated by diagonalizing the matrix $\La$.
In particular, in the case of interest $A=1$, $B=0$ one finds:
\beq
\pmatrix{1\cr 0}(1-CTV)^{-1}=[1+V^{11}CT(T-2)]^{-1}\pmatrix{1\cr T}.
\eeq
We thus obtain the result 
\widetext
\top{-2.8cm}
\hglue -1 cm 
\bea
\pmatrix{V^{12},&V^{21}}(1-CTV)^{-1}CT\pmatrix{V^{21}\cr V^{12}}
&=&
[1+V^{11}CT(T-2)]^{-1}CT[\pmatrix{V^{21},&V^{12}}
+T\pmatrix{V^{12},&V^{21}}]\pmatrix{V^{21}\cr V^{12}}
\nonumber\\
&=&
[1+V^{11}CT(T-2)]^{-1}CT[2(V^{11})^2-2V^{11}C+T(1-(V^{11})^2)] .
\label{expect}
\eea
\bottom{-2.7cm}
\narrowtext
\noindent  
Here,
we used the identities 
\bea
(V^{21})^2=(V^{12})^2&=&(V^{11})^2-V^{11}C ,
\nonumber\\
V^{12}V^{21}+V^{21}V^{12}&=&1-(V^{11})^2 ,
\eea
which can be proved 
via the explicit expressions in terms of $U$ and $C$.
Note that the result in Eq.\ \rf{expect}
involves only $T$, $C$ and $V^{11}$.

We can use the result \rf{expect} in Eq.\ \rf{eqS11} to obtain
an equation for $T$.
Some rearrangement of terms yields
\beq
(T-1)[T^2-(1+X^{-1})T+1]=0,
\label{eqt}
\eeq
where $X\equiv CV^{11}$.
This generates three solutions for $T$:
\beq
T=1,\quad T_\pm={1\over2X}\left(1+X\pm\sqrt{(1+3X)(1-X)}\right),
\label{t}
\eeq
validating the ansatz \rf{ansa}. 
The $T=1$ solution gives $S=C$,
and the corresponding string field is the identity functional
restricted to the bosonic sector.
This solution is expected,
and its appearance serves as a nice check on the formalism.
However,
it is physically irrelevant because it is unnormalizable.

The other two solutions are real for eigenvalues $x$ of $X$
satisfying $-\frac 13\le x\le1$.
The solution $T_-$ behaves as $X$ for small $x$, 
while $T_+=1/T_-$ behaves as $X^{-1}$.
A numerical check on the eigenvalues 
of $X=C V^{11}$ shows they are smaller than one
and converge toward zero,
with the three largest being $0.21$, $0.09$, $0.03$. 
Since the presence of any eigenvalue of $T$
with absolute value greater than or equal to one 
leads to an unnormalizable state,
we consider only the solution $T_-$.

In the ghost sector,
a similar construction can be performed.
Starting with a general squeezed state defined by the matrix $\ti S$
and requiring that its width be invariant under the star product
generates the condition
\beq
-C \ti S C= \pmatrix{\ti V^{12},&\ti V^{21}}
(1-\ti \Si \ti V)^{-1}\ti \Si \pmatrix{\ti V^{21}\cr \ti V^{12}}
+\ti V^{11} ,
\label{eqtildeS11}
\eeq
where
\beq
\ti \Si^{rs}=\pmatrix{\ti S&0\cr 0&\ti S} .
\eeq
Note the minus sign on the left-hand side
of Eq.\ \rf{eqtildeS11}, 
appearing because conjugation 
using the two-vertex introduces a relative minus sign
for the $b$ and $c$ oscillators.

The solution of Eq.\ \rf{eqtildeS11} is completely analogous to
the solution of Eq.\ \rf{eqS11}.
This is because the matrices 
$\ti V^{rs}$ are defined through matrices $C$ and $\ti U$
as in the bosonic case, 
and so all identities leading to 
Eqs.\ \rf{eqt} and \rf{t} have ghost-sector equivalents.

Writing
$\ti S= C \ti T $ and $\ti X = C \ti V^{11}$,
one obtains
\beq
(\ti T+1)[\ti T^2+(\ti X^{-1}-3)\ti T+1]=0 .
\label{eqtildet}
\eeq
This has solutions
\beq
\ti T=-1 ,\quad
\ti T_\pm={1\over2\ti X}\left(3\ti X-1\pm\sqrt{(1-\ti X)(1-5\ti X)}\right).
\label{tildet}
\eeq
The expected piece of the  
unnormalizable identity functional is again found as a solution,
corresponding to $\ti T=-1$.

The two solutions $\ti T_\pm$ are real for all $\ti X$ except for
eigenvalues $\ti x$ in the range
$\frac 15<\ti x<1$.
For small $\ti x$,  
$\ti T_+\approx-\ti X$ is small
while $\ti T_-\approx\ti X^{-1}$ is large.
Since the eigenvalues of $\ti V^{11}$ converge toward zero,
only $\ti T_+$ can lead to a normalizable solution to the ghost sector
of Eq.\ \rf{cSdef}. 
One might wonder whether some of the eigenvalues 
$\ti x$ of $\ti X$ lie in the
interval $\frac 15<\ti x<1$, 
which would lead to complex eigenvalues for $\ti T_+$.
However, 
all the $\ti x$ are negative, 
with the largest three eigenvalues being $-0.66$, $-0.25$, and $-0.08$.

Combining the above results provides 
the squeezed state obeying Eq.\ \rf{cSdef} as
\bea
\ket{\cS} 
&=& \hbox{Det}(1-S^2)^{1/4}
\hbox{Det}(1-\ti S^2)^{-1/2}
\nonumber\\
&&
\qquad \times\exp(\half a^\da S a^\da)
\exp(b^\da E \ti S E^{-1}c^\da )\ket{+} .
\label{cSsol}
\eea
By construction,
it is an element of the star subalgebra of string fields.
A family of `wedge' states also lying in this subalgebra
was recently introduced in Ref.\ \cite{rz}.
It would be interesting to determine explicitly 
the relationship of $\cS$ to these wedge states.

\section{Conversion to the squeezed basis}

Since the star product is idempotent on the string field $\cS$,
one can expect the form of the vertex function
to be substantially simplified 
in a new oscillator basis chosen such that the squeezed
state $\ket{\cS}$ plays the role of the ground state. 
As before,
we begin our considerations in the bosonic sector
and subsequently extend them to the ghosts.

Observing that 
$(a-S\ad) \exp(\half \ad S \ad)\rvac=0$,
we are motivated to define
a new annihilation operator $s$ 
by the Bogoliubov transformation
\beq
s=w(a-S\ad) , \quad 
a=w(s+S\sd) ,
\eeq
where
\beq
w =(1-S^2)^{-1/2}.
\eeq
The operator $s$ annihilates $\ket{\cS}$, 
so the latter can be identified
ae the vacuum $\rvac_s$ in the $s$-oscillator basis:
\bea
\rvac_s&=&\hbox{Det}(w)^{-1/2}
\exp(\half\ad S\ad)\rvac_a,
\nonumber\\
\rvac_a&=&\hbox{Det}(w)^{-1/2}
\exp( -\half\sd S\sd)\rvac_s.
\eea

Applying this transformation to the bosonic sector of the three-vertex, 
we find 
\widetext
\top{-2.8cm}
\hglue -1 cm 
\bea
|V_3\rangle
&=&\exp(\half a^{r\da} V^{rs} a^{s\da})\rvac_{a,123}
\nonumber\\
&\propto&\exp(\half a^{r\da} (V^{rs}-S\de^{rs}) a^{s\da})\rvac_{s,123}
\nonumber\\
&\propto&\exp\left[(\sd+s S)^r\{w(V-S)w \}^{rs}
(\sd+Ss)^s\right]\rvac_{s,123}.
\label{3vertexs}
\eea
In the last line of this equation and in what follows,
it is understood that the symbol $S$ 
is to be interpreted as $S_{mn}\de^{rs}$.
In Appendix A,
the useful identity 
\beq
\exp(\ad A\ad+\ad C a+a B a)=
\hbox{Det}\left[(1-C)e^C\right]^{-1/2}
\exp[\ad(1-C)^{-1}A\ad]
\exp[-\ad\ln(1-C) a]
\exp[a B(1-C)^{-1} a]
\label{bosonicr}
\eeq
is shown to hold for multidimensional oscillators,
where the matrices $A$, $B$, and $C$ satisfy
\beq
A^T=A, \quad 
B^T=B, \quad  
AC^T=CA, \quad 
BC=C^TB, \quad 
C^2=4AB.
\label{conditions}
\eeq
\narrowtext
\noindent  
The identity \rf{bosonicr}
can be used to rewrite Eq.\ \rf{3vertexs},
since the appropriate identifications of $A$, $B$, $C$
satisfy the conditions \rf{conditions}.
Noting that exponentials 
involving the annihilation operator $s$
act as the identity on the vacuum,
we find
\beq
|V_3\rangle\propto\exp (\half s^{r\da}\hat V{}^{rs} s^{s\da})
\rvac_{s,123},
\eeq
where
\beq
\hat V =(1-VS)^{-1}(V-S)
\eeq
is the transformed three-vertex function.

Next,
we investigate the properties of $\hat V$.
Using the explicit form \rf{expressV},
$V^{rs}$ can be diagonalized in the $r$, $s$ indices
by a matrix $\cO$ satisfying ${\cO}^{-1}={\cO}^\da$:
\beq
V={\cO}^{-1}V_D{\cO} ,
\eeq
with 
\beq
V_D=\pmatrix{C&0&0\cr 0&U&0\cr 0&0&\bar U\cr} , \quad
{\cO}={1\over\sqrt3}\pmatrix{1&1&1\cr\al^*&\al&1\cr\al&\al^*&1\cr} .
\eeq
For the transformed three-vertex $\hat V$,
we find
\beq
\hat V ={\cO}^{-1}(1-V_DS)^{-1}(V_D-S){\cO},
\eeq
with the $rr$ elements of the diagonalized form given by
\bea
((1-V_DS)^{-1}(V_D-S))^{11}
&=&(1-CS)^{-1}(C-S)
\nonumber\\
&=&C,
\nonumber\\
((1-V_DS)^{-1}(V_D-S))^{22}
&=&(1-US)^{-1}(U-S),
\nonumber\\
((1-V_DS)^{-1}(V_D-S))^{33}
&=&(1-\bar US)^{-1}(\bar U-S).
\label{transformedev3}
\eea

The diagonal elements of
the transformed vertex function $\hat V^{rs}$
are all equal and given as one-third of the sum 
of its eigenvalues in Eqs.\ \rf{transformedev3}, 
in analogy with $V^{rs}$.
Multiplying any one diagonal element $\hat V^{rr}$
from the left and from the right
with $(1-US)$ gives $V^{11}(1+S^2)-V^{11}CS-S$,
which vanishes for the solution $S=CT_-$ in Eq.\ \rf{t}.
This is to be expected, 
since the $s$-vacuum can satisfy
$\rvac_s\star\rvac_s=\rvac_s$ 
only if indeed the diagonal elements of $\hat V$ vanish.

The off-diagonal elements of $\hat V$ are also of interest.
Explicitly,
we find 
\beq
\hat V=\pmatrix{0&\hat V^{12}&\hat V^{21}\cr
\hat V^{21}&0&\hat V^{12}\cr
\hat V^{12}&\hat V^{21}&0\cr},
\eeq
where
\bea
\hat V^{12}&=&\frac 13 (C+\al\La+\al^*\bar\La) ,\nonumber\\
\hat V^{21}&=&\frac 13 (C+\al^*\La+\al\bar\La) ,\nonumber\\
\La&=&(1-US)^{-1}(U-S) ,\nonumber\\
\bar\La&=&(1-\bar US)^{-1}(\bar U-S)=C\La C.
\eea
It can be shown that
\beq
C+\La+\bar\La=0 , \quad 
\La^2=\bar\La^2=1.
\label{La2one}
\eeq
Using these results,
some straightforward algebra yields 
\beq
(\hat V^{12}\hat V^{21})^2=\hat V^{12}\hat V^{21} , \quad 
(\hat V^{21}\hat V^{12})^2=\hat V^{21}\hat V^{12},
\label{P12}
\eeq
demonstrating that 
\beq
P_1\equiv\hat V^{12}\hat V^{21} , \quad 
P_2\equiv\hat V^{21}\hat V^{12}
\eeq
are projection operators.
It also follows that $P_1+P_2=1$,
which implies $P_1$ and $P_2$ are conjugate.
Other useful identities are
\bea
P_1 &=& \hat V^{12} C = C \hat V^{21},
\quad 
P_2 = \hat V^{21} C = C \hat V^{12},
\nonumber\\
\hat V^{12}&=&\hat V^{12}P_2=P_1\hat V^{12},
\quad
\hat V^{21}=\hat V^{21}P_1=P_2\hat V^{21},
\nonumber\\
&&\qquad\quad  (\hat V^{12})^2=(\hat V^{21})^2=0.
\label{hatV2zero}
\eea

We interpret the physical meaning of
Eqs.\ \rf{P12} through \rf{hatV2zero}
in terms of the behavior of 
the left half (``1'') and the right half (``2'')
of the string.
Thus, 
$\hat V^{12}$ maps the right half onto the left half
and annihilates the left half,
while $\hat V^{21}$ does the converse.
The operators $P_1$ and $P_2$
project onto the left and right halves
of the string, respectively.
This interpretation is consistent with the identity
$\hat V^{12}+\hat V^{21}=C$,
since the operator $C$ indeed interchanges
the left and right halves of the string.
The existence of such an interpretation is to be expected
because the action of the three-vertex is to map the left half of string $r$ 
onto the right half of string $r+1$ modulo 3.
The Bogoliubov transformation to the $s$-oscillator basis
apparently generates a structure reminiscent 
of the `comma' representation proposed in Ref.\ \cite{bcnt}.
It would be interesting to obtain explicitly 
the relation between the two formulations.

Turning next to the ghost sector,
we define new fermionic ghost oscillators $t$ and $u$ 
through the Bogoliubov transformations
\bea
t&=&(b E-\bd E\ti S)\ti w E^{-1},
\quad 
u=E\ti w(E^{-1}c-\ti SE^{-1}\cd),
\nonumber\\
b&=&(t E+\td E S)\ti w E^{-1},
\quad
c=E\ti w(E^{-1}u+SE^{-1}\ud),
\nonumber\\
&&\qquad\qquad
\ti w=(1-\ti S^2)^{-1/2}.
\eea
The oscillators $t$, $u$ satisfy
the same algebra as the $b$, $c$ oscillators. 
Also,
$t$, $u$ annihilate the state $\exp(\bd\ti S\cd)\ket{+}_{b,c}$,
so we identify the vacuum $\ket{+}_{t,u}$ as
\bea
\ket{+}_{t,u}&=&\hbox{Det}(\ti w)
\exp(\bd E\ti SE^{-1}\cd)\ket{+}_{b,c},\nonumber\\
\ket{+}_{b,c}&=&\hbox{Det}(\ti w)
\exp(-\td E\ti SE^{-1}\ud)\ket{+}_{t,u}.
\eea

\widetext
The ghost three-vertex can be expressed in terms of the new oscillators:
\beq
|V_3^{gh}\rangle\propto\exp\left[(\td E-t E\ti S)^r
[\ti w (\ti V-\ti S)\ti w]^{rs}
(E^{-1}\ud+SE^{-1}u)^s\right]
\ket{+}_{t,u,123}.
\label{3vertextu}
\eeq
We find that the ghost version of the identity \rf{bosonicr} 
for multidimensional oscillators is
\bea
\exp(\bd A\cd+\bd C c+\cd C b+b B c)&=&
\hbox{Det}\left[(1-C)e^C\right]
\exp[\bd(1-C)^{-1}A\cd]
\nonumber\\
&&\qquad\qquad
\times\exp[-\bd\ln(1-C) c-\cd\ln(1-C) b]
\exp[b B(1-C)^{-1} c],
\label{ghostriccati}
\eea
where the matrices $A$, $B$, and $C$ satisfy
\beq
A^T=A , \qquad
B^T=B , \qquad
AC^T=CA , \qquad
BC=C^TB , \qquad
C^2=-AB .
\label{ghostconditions}
\eeq
Using the identity \rf{ghostriccati} in Eq.\ \rf{3vertextu}
yields
\beq
|V_3^{gh}\rangle\propto\exp[t^{\da} E(1-\ti V\ti S)^{-1}
(\ti V-\ti S)E^{-1} u^{\da}]
\ket{+}_{t,u,123}.
\eeq
\bottom{-2.7cm}
\narrowtext
\noindent  

In analogy with the bosonic sector,
it can be shown that the transformed ghost vertex
\beq \hat{\ti V}=
(1-\ti V\ti S)^{-1}(\ti V-\ti S)
\eeq
has zero diagonal $rr$ elements.
Its off-diagonal elements can be used to define projection operators 
\beq
\ti P_1=\hat{\ti V}{}^{12}\hat {\ti V}{}^{21} , \quad
\ti P_2=\hat{\ti V}{}^{21}\hat{\ti V}{}^{12}.
\eeq
The interpretation in terms of left and right halves of the string
also holds here.

\section{The nonperturbative vacuum $\cN$}

\subsection{Construction}

We next consider the equations of motion for the string field
and construct a solution for the nonperturbative vacuum $\cN$.
To simplify the notation in expansions of string fields,  
we use Greek indices to indicate a composite index
containing both level-number indices and spacetime Lorentz indices.
In the squeezed-state basis,
a general string field $\ket f$ can then be expanded as
\widetext
\top{-2.8cm}
\hglue -1 cm 
\beq
\ket f=\sum_{m,n=0}^\infty 
f_{m,n}^{\la_1\ldots\la_m\mu_1\nu_1\ldots\mu_n\nu_n}
s_{\la_1}^\da \ldots s_{\la_m}^\da t_{\mu_1}^\da
u_{\nu_1}^\da\ldots t_{\mu_n}^\da u_{\nu_n}^\da
\ket{-}_{s,t,u}.
\label{expand}
\eeq
\bottom{-2.7cm}
\narrowtext
\noindent  
Here,
$f_{m,n}$ is a tensor 
that is totally symmetric in $\la_1\ldots\la_m$
and totally antisymmetric in $\mu_1\ldots\mu_n$
and in $\nu_1\ldots\nu_n$. 
The indices $m$, $n$ here label the number of 
composite bosonic and ghost indices
rather than oscillator numbers,
the latter being subsumed into Greek indices 
as explained above.

We are interested in solving the string equation of motion
$Q\Ps +g\ap\Ps \star\Ps =0$, 
which for $g = \ap = 1$ reduces in the Feynman-Siegel gauge 
\cite{ws}
to
\beq
c_0(L_0-1)\Ps+\Ps\star\Ps=0.
\label{sfteqsf}
\eeq
Restricting attention first to the bosonic sector,
the action of $L_0$ in the $s$-oscillator basis is found to be
\widetext
\top{-2.8cm}
\hglue -1 cm 
\bea
L_0|f\rangle&=&\sum_{m=0}^\infty \Bigl[
(\sd wSE^{-2}w\sd)
f_{m0}^{\la_1\ldots\la_m} s_{\la_1}^\da\ldots s_{\la_m}^\da
+m(m-1)(wSE^{-2}w)_{\la_{m-1}\la_m}
f_{m0}^{\la_1\ldots\la_m} s_{\la_1}^\da \ldots s_{\la_{m-2}}^\da
\nonumber\\
&&\qquad\qquad {}
+m(\sd w(E^{-2}+SE^{-2}S)w)_{\la_m}
f_{m0}^{\la_1\ldots\la_m} s_{\la_1}^\da\ldots s_{\la_{m-1}}^\da
+\De f_{m0}^{\la_1\ldots\la_m} s_{\la_1}^\da \ldots s_{\la_m}^\da
\Bigr]\rvac_s.
\label{L0}
\eea
Here,
$\De$ is the energy 
of the squeezed $s,t,u$ vacuum relative to the usual $a,b,c$ vacuum,
as measured by the action of $L_0$
and arising from normal ordering in the squeezed vacuum. 
It is given by
\beq
\De=26\De^s+\De^{t,u} , \quad 
\De^s=\hbox{Tr}(wSE^{-2}Sw)
=\hbox{Tr}[S^2 (1-S^2)^{-1}E^{-2}] , \quad 
\De^{t,u}=-2\hbox{Tr}[\tilde S^2(1-\tilde S^2)^{-1}E^{-2}].
\label{Des}
\eeq
The star product of two states $\ket f$ and $\ket g$
restricted to the bosonic sector is 
\bea
|f\star g\rangle&=&\sum_{m,n}\sum_{k=0}^{\hbox{min}(m,n)}
[(\hat V^{21})_{\la_1\mu_1}\ldots(\hat V^{21})_{\la_k\mu_k}]
[(P_1)_{\la_{k+1}}^{\pt{\la_{k+1}}\la'_{k+1}}\ldots
(P_1)_{\la_m}^{\pt{\la_m}\la'_m}]
\nonumber\\
&&\qquad\qquad\qquad\times
[(P_2)_{\mu_{k+1}}^{\pt{\mu_{k+1}}\mu'_{k+1}}\ldots
(P_2)_{\mu_n}^{\pt{\mu_n}\mu'_n}]
f_{m0}^{\la_1\ldots\la_m}
g_{n0}^{\mu_1\ldots\mu_n} 
s_{\la'_{k+1}}^\da \ldots s_{\la'_m}^\da
s_{\mu'_{k+1}}^\da \ldots s_{\mu'_n}^\da\rvac_s.
\label{fstarg}
\eea
\narrowtext

Define the excitation number $m$ of a term in the field expansion
as the associated number of $\sd$ factors acting on the $s$ vacuum.
Then,
Eq.\ \rf{L0} reveals that 
$L_0$ maps excitation number $m$ 
to excitation numbers $m-2$, $m$, and $m+2$,
while Eq.\ \rf{fstarg} shows that 
the star operator connects excitation numbers $m$ and $n$
to a combination of excitation numbers
$m+n$, $m+n-2$, $\ldots$, $|m-n|$. 
The latter is reminiscent of the combination of two angular momenta.
This is no accident,
as we demonstrate next.

In the $s$-oscillator basis,
the existence of the projection operators $P_1$ and $P_2$
separates the Greek indices 
into degrees of freedom pertaining to left and right halves of the string.
This split enables the realization of an SU(2) symmetry
associated with excitation number.
Consider the operators 
\bea
J_0&=&\half(s^\dagger P_1 s - s^\dagger P_2 s),
\nonumber\\
J_+&=&s^\dagger \hat V^{12} s , \quad
J_-=s^\dagger \hat V^{21} s.
\eea
These form the generators of the excitation su(2) algebra.
Note that the $J_\pm$ connect left and right oscillators.
The total spin $J = s^\dagger s/2$ characterizes the representations.
This means that states of excitation number $n$ 
constitute a representation of the excitation SU(2) symmetry 
with spin $J=n/2$.

In general,
the star product between states 
of excitation numbers $m$ and $n$
contains only min$(m,n)+1$ nonzero states of the $mn$ possibilities,
consisting of states of excitation $|m-n|$ through $m+n$.
As an example,
Eq.\ \rf{fstarg} shows 
the star product of two $J=1/2$ representations
$\ket{f}_s \equiv f_{10}^\la s_\la^\da\rvac_s$
and 
$\ket{g}_s \equiv g_{10}^\la s_\la^\da\rvac_s$
is
\beq
\ket{f\star g}=
f_{10}^\la g_{10}^\mu 
[(\hat V{}^{21})_{\la\mu} + (P_1 s^\da)_\la (P_2 s^\da)_\mu ]\rvac_s .
\eeq
Two nonzero states remain,
one of excitation 0 and one of excitation 2.

The above discussion suggests it may be useful to distinguish 
even and odd excitation numbers. 
In particular,
the action of both $L_0$ and the star operator 
are closed in the subsector of even excitation number.
In the remainder of this section, 
we restrict attention to this subsector. 

Despite the simplifications offered by the squeezed basis,
obtaining an explicit exact solution to Eq.\ \rf{sfteqsf}
remains a somewhat formidable task
given the complexity of Eqs.\ \rf{L0} and \rf{fstarg}.
The approach we adopt here relies on the observation
that the matrix $S$ appears in the definition of $L_0$.
Following the discussion in section III,
it can be shown that the eigenvalues of $S$ are small.
The largest eigenvalues are approximately 
$-0.21$, $0.09$, and $-0.03$, 
displaying an alternating series converging to zero.
It is therefore a reasonable strategy to 
solve Eq.\ \rf{sfteqsf} perturbatively in $S$.

We expand the desired string field $\cN$ as 
\beq
\ket \cN =\ket{\cN^{(0)}}+\ket{\cN^{(1)}}+\ket{\cN^{(2)}}+\ldots,
\eeq
where the superscript indicates the order in $S$. 
Substitution in Eq.\ \rf{sfteqsf} 
at order $S^0$ gives
\bea
&&\sum_{m=0}^\infty (m s^\dagger E^{-2}-s^\dagger)_{\la_1}
s^\dagger_{\la_2}\ldots s^\dagger_{\la_m}
{\cN}_{m0}^{\la_1\ldots\la_m(0)} \rvac_s
\nonumber\\
&& \qquad \qquad 
+\ket{{\cN}^{(0)}}\star\ket{{\cN}^{(0)}}=0.
\label{orderzero}
\eea
Note that the energy of the squeezed vacuum $\De\sim\cO (S^2)$ 
is irrelevant at this order.
One solution is obtained by setting to zero
all terms with $m\ne0$,
yielding
\beq
\ket{\cN^{(0)}}+\ket{\cN^{(0)}}\star\ket{\cN^{(0)}}=0.
\label{order0}
\eeq
We can use the analysis in section IV to set 
\beq
\ket{\cN^{(0)}} = \ket \cS\equiv \rvac_s,
\eeq
so that
\beq
\cN_{00}^{(0)}=1, 
\quad 
\cN_{20}^{\la_1\la_2}{}^{(0)}=\cN_{40}^{\la_1\ldots\la_4}{}^{(0)}
=\ldots=0.
\eeq
There may also be other possible choices
producing different solutions to the string equations of motion.
We return to this issue in the next section.

At order $S^1$, 
one obtains
\bea
&&[(\sd SE^{-2}\sd)\cN_{00}^{(0)} 
\nonumber\\
&&\qquad{}
+\sum_m (m\sd E^{-2}-\sd)_{\la_1}
s_{\la_2}^\da \ldots s_{\la_m}^\da \cN_{m0}^{\la_1\ldots\la_m}{}^{(1)}]
\rvac_s
\nonumber\\
&&\qquad{}
+\ket{\cN^{(1)}}\star\ket{\cN^{(0)}}
+\ket{\cN^{(0)}}\star\ket{\cN^{(1)}}
=0,
\label{orderep}
\eea
where the star product is understood to be expanded in components
at the appropriate order,
according to Eq.\ \rf{fstarg}.
Note that the star product of 
$\ket{\cN^{(0)}}=\rvac_s$ with
a term in $\cN^{(1)}$ of excitation number $m$ 
yields a result with the same excitation number $m$.
The presence of the first (inhomogeneous) term
in Eq.\ \rf{orderep} means that
the only nonvanishing contribution at order $S$
involves excitation number 2.
In particular,
$\cN_{00}^{(1)} = 0$.

The explicit solution of Eq.\ \rf{orderep} for $\ket{\cN^{(1)}}$ 
yields a somewhat cumbersome expression, 
involving left and right projection operators.
However, 
the solution at order $S$
simplifies greatly in the limit of high mass level,
because the single term involving $mE^{-2}$ dominates 
all terms other than the first. 
In this limit, 
we find
\beq
\cN_{20}^{\la_1\la_2}{}^{(1)}\approx -\half S^{\la_1\la_2}.
\label{orderepf2}
\eeq
Reconverting to the $a$-oscillator basis gives
\bea
\ket \cN &\approx&(1-\half\sd S\sd)\rvac_s
\nonumber\\
&\approx&(1-\half\ad S\ad)(1+\half\ad S\ad)\rvac_a +{\cO}(S^2)
\nonumber\\
&=&\rvac_a +{\cO}(S^2).
\eea
This shows that at high mass levels
the field $\cS$ dominating the lowest-order solution
is cancelled.
This is consistent with the expected structure 
of the nonperturbative vacuum $\cN$,
with low-mass levels taking values near the squeezed field $\cS$
and high-mass levels taking ones close to the usual vacuum.

The contributions to excitation number two arising from
the last star-product term in Eq.\ \rf{orderep}
consist of $P_2$ projections on each index of
$\cN_{20}^{\la_1\la_2}{}^{(1)}$.
Similarly,
the preceding term involves $P_1$ projections.
Moreover,
the contributions of unmixed projections in 
the term involving $-\sd$ cancel,
so the result \rf{orderepf2}
in fact holds exactly at order $S$
for all mass-level contributions 
involving unmixed projections.
Also,
at order $S$ the mixed projections are absent from
the final two star-product terms.
However,
contributions from the $-\sd$ term do arise for the mixed projections.
Schematically,
one finds in this case that the result \rf{orderepf2}
for $\cN_{20}^{(1)}$
becomes replaced by a structure of the general form 
$-SE^{-2}/(2E^{-2}-1)$. 
For the subleading mass level this produces a contribution
$-S$,
thereby changing the sign of the term proportional to $S$
in the lowest-order approximation.
Since the true state acquires a combination of contributions
from mixed and unmixed projections,
one again can anticipate an effect intermediate 
between $\cS$ and the usual vacuum at the lowest mass levels.
We see that the nonperturbative vacuum $\cN$
can be regarded as a twisted squeezed state,
constructed by the action of operators on the squeezed field $\cS$
and with net field values dropping more rapidly with excitation number 
than an exponential at this order of approximation.

The order-$S$ solution can now be used to advance to order $S^2$.
A nonzero contribution to $\cN_{40}^{(2)}$ arises at this level,
along with corrections to the lower-order results.
This pattern continues at higher orders,
with the first contribution to $\cN_{2N,0}$ being at order $S^N$.
It would be interesting to determine whether 
a mechanism analogous to that described above
causes higher-order expectation values 
also to approach zero faster than an exponential 
as the excitation level increases.

A complete treatment at order $S^2$
requires developing a method to handle
divergent traces that appear in some terms.
For example,
a contribution arises to the vacuum-energy shift $\De$ 
in Eq.\ \rf{Des} that is proportional to Tr$(S^2E^{-2})$
and that appears to have a linear divergence.
It is possible that combining this with the ghost contribution
would yield a finite result,
but in general it may be necessary to regulate such terms.
Zeta-function regularization may be most appropriate,
since it is known to avoid an associativity anomaly in the vertex 
in related calculations 
\cite{pt}.

In the ghost sector,
an analogous construction 
for the nonperturbative vacuum $\cN$ can be performed.
It is straightforward to extend the formal analysis to 
a solution in powers of $\ti S$
and to extract results at order $\ti S$.
The order $\ti S^0$ equation has the same form as Eq.\ \rf{order0},
so the complete lowest-order solution
can be taken as $\cN_{00}^{(0)}=\ket{-}_{s,t,u}$
with all other components zero.
The equivalent of Eq.\ \rf{orderep} for the ghost sector is 
\bea
&&[
(\td E\{\ti S,E^{-2}\} E^{-1}\ud)\cN_{00}^{(0)} 
\nonumber\\
&&\quad\qquad
+[(\td E^{-2})_{\mu_1} u_{\nu_1}^\da
+t_{\mu_1}^\da ((\ud E^{-2})_{\nu_1})]
\cN_{02}^{\mu_1\nu_1}{}^{(1)}
\nonumber\\
&&\quad\qquad
+ \hbox{higher-level terms}]\ket{+}_{t,u}
\nonumber\\
&&\quad
+\ket{\cN^{(1)}}\star\ket{\cN^{(0)}}
+\ket{\cN^{(0)}}\star\ket{\cN^{(1)}}
=0.
\label{orderepgh}
\eea
Contributions from excitation number greater than 2
and from $\cN_{00}^{(1)}$
again can be neglected at this order.

The analogue of Eq.\ \rf{orderepf2}
in the limit of high mass level is
\beq
\cN_{02}^{\mu_1\nu_1}{}^{(1)}\approx 
-(E\ti SE^{-1})^{\mu_1\nu_1}.
\label{orderepf2gh}
\eeq
Evidently,
the contributions from ghost fields at excitation number 2
become of order $\ti S^2$ in this limit.
At low mass levels,
an intermediate effect between cancellation 
of the term linear in $\ti S$ and a reversal of the sign
can again be expected.

\subsection{Properties}

Since the form of our solution is Lorentz invariant by construction,
we expect $\cN$ to correspond to the Lorentz-invariant 
nonperturbative vacuum of Refs.\ \cite{ksobs,kspc,kp2}.
The results in the previous subsection
can be used to compare the solution for $\cN$
with numerical values for nonperturbative solutions 
obtained in the level-truncation scheme.

In Table 1 we list numerical approximations for 
some component bosonic-sector scalar fields 
in the Lorent-invariant nonperturbative vacuum
obtained \cite{mt}
at level truncation $(10,20)$,
together with the range in which 
the $\cO(\cS)$ solution for $\cN$
indicates the values should lie.
The discussion in the previous subsection
implies that the true values are expected 
to lie near the maximum of this range for low-lying states 
but to fall to the minimum of the range as the level number increases.
These results support the identification of $\cN$
with the level-truncated Lorentz-invariant nonperturbative vacuum.

In the ghost sector,
the numerical values for the ghost-oscillator fields
follow a pattern similar to that in the bosonic sector.
However,
a numerical approximation for $\ti S$ is more involved than
for $S$ because approximating $\ti T_+$ 
by Taylor expansion of the square root in Eq.\ \rf{tildet}
fails for the largest two eigenvalues of $\ti X$,
which lie outside the radius of convergence.  
It would interesting to find a means of approximating
$\ti S$ with sufficient accuracy 
to make possible comparisons with truncation-scheme calculations.

At order $\ti S^2$,
the level of technical complication increases
and the corresponding analysis lies beyond the scope of this work.
We conjecture that the cancellation mechanism
occurring at linear order in the limit of high mass level generalizes
at higher order.
This is consistent with the pattern emerging from 
numerical approximations using the truncation scheme.

\begin{center}
\begin{tabular}{|c|c|c|}
\hline
{} \qquad State \qquad {} & {}\qquad $\langle\ps\rangle$ \qquad {} &
{} \quad Expected Range \quad {} \\
\hline
\hline
$\rvac$ & 1.09259 & 1 \\
\hline
$\ad_1\cdot\ad_1\rvac$ & 0.05723 & 0 to 0.069 \\
\hline
$\ad_1\cdot\ad_3\rvac$ & -0.01988 & -0.053 to 0 \\
$\ad_2\cdot\ad_2\rvac$ & -0.01018 & -0.023 to 0 \\
\hline
$\ad_1\cdot\ad_5\rvac$ & 0.00783 & 0 to 0.032 \\
$\ad_2\cdot\ad_4\rvac$ & 0.00823 & 0 to 0.031 \\
$\ad_3\cdot\ad_3\rvac$ & 0.00429 & 0 to 0.017 \\
\hline
\end{tabular}
\end{center}
\noindent
{\small
Table 1. Vacuum expectation values for bosonic-sector
scalar states in the nonperturbative vacuum.
For each state, 
the expectation value $\langle\ps\rangle$ 
evaluated in the level-truncation scheme 
and the range allowed by the order-$S$ calculation
in the text are presented.
The correct value is predicted to be near 
the maximum of this range for low-lying states 
but rapidly to approach the minimum 
as the level number increases,
in agreement with the results from 
the level-truncation scheme.}
\medskip

It is of interest to compare the vacuum energy 
$E_{\cN}$ of $\cN$ to the D25-brane mass $M_{25}$.
In principle,
it suffices to evaluate the action $I(\Ps)$ 
for the on-shell string field $\Ps = \cN$
and use $E_{\cN} = I(\cN)$.
In the present context,
the evaluation can be performed directly at order $S$, $\ti S$:
\bea
E_{\cN}= I(\cN)
&=& \frac 12 \int \cN\star Q \cN
+ \frac 13 \int \cN \star \cN \star \cN
\nonumber\\
&=& -\frac 16 \int \cN \star \cN \star \cN
\nonumber\\
&\approx & -\frac 16 \int \cS \star \cS \star \cS
\nonumber\\
&=& - \frac 16 
{~}_{123,s,t,u}\bra{-}
V_3\rangle_{s,t,u}
\nonumber\\
&\approx &-\frac 16 + \cO (S^2, \ti S^2)
\nonumber\\
&\simeq& -0.17 .
\label{evalact}
\eea
This derivation takes advantage of 
the vanishing of the diagonal elements
of the 3-vertex in the $s$, $t$, $u$, basis,
proved in section IV. 
The result is to be compared with the expected value 
$E_{\cN} = - M_{25}= - 2/\pi^2\approx -0.20$.
Approximating the field $\cN$ by $\cS$
thus gives about 85\% of the D-brane mass.

The string action can also provide insight 
into the nature of excitations about the squeezed field $\cS$.
Consider first a string field $\Ps$
expanded about an arbitrary background $\Ps_B$:
$\Ps= \Ps_B + \De$.
The action for $\Ps$ can be written 
\cite{kstusc}
\bea
I(\Ps) &=& 
\fr 1 {2\ap} \int \Ps_B \star Q\Ps_B
+ \fr g 3 \int \Ps_B \star \Ps_B \star \Ps_B
\nonumber\\
&&\quad
+ \fr 1 {2\ap} \int \De * Q_B\De
+ \fr g 3 \int \De\star\De\star\De .
\label{action}
\eea
Here,
the action of the background BRST operator $Q_B$ 
on an arbitrary string field $\Ph$ 
of ghost number $g(\Ph)$ is given by
\beq
Q_B \Ph = Q \Ph 
+ g\ap [\Ps_B \star \Ph - (-1)^{g(\Ph)}\Ph\star\Ps_B] .
\label{qb}
\eeq
The background operator $Q_B$ is nilpotent and distributive
across the star product:
$Q_B(\Ph_1\star\Ph_2) 
= Q_B\Ph_1\star\Ph_2 
+ (-1)^{g(\Ph_1)}\Ph_1\star Q_B\Ph_2$.
It also satisfies
$\int Q_B\Ph_1\star\Ph_2 
= (-1)^{g(\Ph_1)+1}\int\Ph_1\star Q_B\Ph_2$.
The action $I(\Ps)$ is invariant under
\beq
\de \De = \fr 1 {\sqrt{2\ap}} Q_B \La 
+ g \sqrt{\fr{\ap}{2}}
(\De \star \La - \La \star \De),
\label{sym}
\eeq
where $\La$ is a string gauge field. 

In the special case that $\Ps_B = \cS$,
the fluctuations $\De$ represent oscillations
about the squeezed field $\cS$  
and the background BRST operator $Q_B = Q_\cS$
determines the corresponding spectrum.
We conjecture that this operator has no tachyonic modes.
This would provide further support for the idea 
that the squeezed field $\cS$ provides
a useful starting point for investigating
the nonperturbative structure of the open bosonic string,
despite not being a solution of the string equations of motion.

As partial support for the conjecture,
we verify it at order $S$ for low-lying states
in the bosonic sector.
Consider first the state $\rvac_s$.
The $\cS$-background BRST operator $Q_\cS$
acting on this state gives
\bea
b_0 Q_\cS \rvac_s
&\approx &
(s^\da E^{-2}s - 1) \rvac_s
+ 2 \rvac_s \star \rvac_s
\nonumber\\
&\approx &
- \rvac_s + 2 \rvac_s \star \rvac_s = +\rvac_s.
\eea
This approximation suffices to show that,
instead of a tachyonic mass as in the vacuum $\rvac_a$,
the state $\rvac_s$ acquires a conventional mass.

Similar reasoning supports the conjecture
for the first excited states
$s_n^\da\rvac_s$:
\bea
b_0 Q_\cS s_n^\da \rvac_s
&\approx &
(s^\da E^{-2}s - 1) s_n^\da \rvac_s
\nonumber\\
&&\qquad
+ \rvac_s \star s_n^\da \rvac_s + s_n^\da \rvac_s \star \rvac_s
\nonumber\\
&=& (n-1) s_n^\da \rvac_s 
\nonumber\\
&&\qquad
+ P_1 s_n^\da \rvac_s \star \rvac_s
+ \rvac_s \star P_2 s_n^\da \rvac_s 
\nonumber\\
&=& n s_n^\da \rvac_s .
\label{vecmass}
\eea
Since $n>0$,
this calculation shows that the
analogue of the states $a_n^\da \rvac_a$ that contain massless modes
is here a set of purely massive states $s_n^\da \rvac_s$.

It is believed that the endpoint of tachyon condensation
on the D25-brane has no physical open-string excitations
\cite{as}.
The corresponding conjecture in the present context
is that no normalizable states can be constructed on $\cN$.
Equivalently,
setting $\Ps_B = \cN$ in the background equations above,
one expects the background BRST operator $Q_B = Q_\cN$
to have no normalizable physical fluctuations $\De$,
so the coefficients of all physical kinetic terms for $\De$ 
should vanish.
An alternative would be to show there are no zeros in the
euclidean propagators for physical fields 
in the nonperturbative vacuum
\cite{ksobs}.
Although our construction of $\cN$ sets the momenta to zero,
it may be possible to generalize it
to incorporate euclidean propagators for $\De$
(\it cf.\ \rm remarks in the next section).
In any event,
it would be interesting to have an explicit proof 
of this conjecture in the context of string field theory.

We note in passing that states on which the action
of creation and annihilation operators 
produces only unnormalizable results can readily be found.
For example,
in the context of a one-dimensional bosonic oscillator
the state 
$\ket{\ps} = \sum_n [(\ad )^n/n\sqrt{n!}]\rvac$
has finite norm $\sum_n n^{-2} = \pi^2/6$,
but the norms of both 
$\ad \ket{\ps}$ and $a \ket{\ps}$
behave as $\sum_n n^{-1}$ and so diverge.
A multidimensional version of this is the state 
$\Pi_m \sum_n [(a_m^\da )^n/n\sqrt{n!}]\rvac$,
which itself is normalizable but
produces unnormalizable states when acted on by
any of the oscillator creation and annihilation operators
$a_m^\da$, $a_m$.
We anticipate that the nonperturbative vacuum $\cN$
has similar features.

A related issue is the fate of the
low-energy U(1) or U(N) particle symmetry 
of the perturbative massless vectors
in the nonperturbative vacuum $\cN$.
In string field theory,
the nonabelian gauge symmetry 
is modified compared to the usual transformation law.
It has been shown that the nonperturbative vacuum $\cN$
maintains the string U(N) symmetry contained in Eq.\ \rf{sym}
in the sense that nonzero scalar expectation values
remain invariant under a string U(N) transformation 
to all truncation orders
\cite{kspc}.
However,
this low-energy string symmetry is distinct from the 
usual U(N) particle gauge symmetry,
with the two being related through a set of nonlinear field redefinitions
\cite{kpp}.
The effect of this 
on the generation of a mass term for the U(1) field
has recently been studied 
in the context of the level-truncation scheme
\cite{wt2}.
Confinement via the condensation of magnetically charged tachyons
has been suggested
\cite{bhy},
the possibility of a critical value for the U(1) field
above which no solutions exist has been investigated \cite{sz2},
and it has also been proposed that as the tachyon condenses 
the noncommutative gauge symmetry is fully unbroken
and becomes a linearly realized U($\infty$)
forbidding propagation of open-string modes
\cite{gms}.
It would be interesting to investigate  
the U(N) particle symmetry
in the context of the solution $\cN$ found here.
The nonzero mass for vector excitations about 
the squeezed field $\cS$,
demonstrated by Eq.\ \rf{vecmass},
suggests this particle symmetry is destroyed in the 
nonperturbative vacuum.

\section{Discussion}
\label{Discussion}

Using analytical methods,
we have constructed a nonperturbative vacuum $\cN$ 
satisfying the full equations of motion 
\rf{eqmot}
of string field theory.
It appears to coincide with 
the Lorentz-invariant nonperturbative vacuum 
found using the level-truncation scheme
and resulting from tachyon condensation on a D25-brane.

In the original $(a,b,c)$-oscillator basis 
and in the Siegel-Feynman gauge,
the form of $\cN$ is given as
\beq
\ket{\cN} = F(S, \ad) \tilde F(\tilde S, \bd, \cd) \ket{\cS}.
\label{summeq}
\eeq
Here,
the squeezed field $\ket{\cS}$ satisfying Eq.\ \rf{cSdef}
is explicitly given in closed form in Eq.\ \rf{cSsol}.
Closed-form expressions for the matrices 
$S=CT_- $ and $\tilde S = C\tilde T_+$
in terms of the 3-vertex functions $V^{11}$ and $\tilde V^{11}$
in the momentum representation
are determined by Eqs.\ \rf{t} and \rf{tildet}.
The product $F\tilde F\approx 1 + \cO(S, \tilde S)$
is constructed recursively in powers of $S$ and $\tilde S$ 
in section V.

The string field $\cN$ can be regarded as 
a twisted squeezed state.
The construction \rf{summeq} provides $\cN$ in the form
of a series of creation operators
acting on a squeezed field $\cS$.
We suspect a closed form exists for this twist series.
The rapid decay of the expectation values with level number
appears to be a crucial feature of $\cN$.
The occurrence of the operator $L_0$ in the equations of motion 
and its growth with level number
makes it a natural candidate for controlling
this rapid decay and hence the twist series,
and we therefore conjecture that the closed form for $\cN$
is a twisted squeezed state
with twist explicitly determined by $L_0$. 
In any case,
it would be of interest to develop systematic methods
for handling the twist series.
Note also that the squeezed field $\cS$ itself
may merit further study in its own right.
Although $\cS$ does not solve the equations of motion,
it may be a better starting point 
than the usual perturbative vacuum
for approximation schemes studying the properties of $\cN$.

An interesting issue is whether 
solutions exist to Eq.\ \rf{orderzero}
other than $\cN$.
This would imply that at least some terms 
with $m\ne0$ in Eq.\ \rf{orderzero} are nonzero.
Since the star product between states  
with excitation numbers $m_1$ and $m_2$ 
generally yields terms up to excitation number $(m_1+m_2)$, 
it seems likely that nonzero terms persist for 
arbitrarily high excitation number.
Any other solution to Eq.\ \rf{orderzero}
is therefore unlikely to be simple within this framework
even at low order.
Also,
satisfying the criterion of normalizability may be more difficult for 
any such solution because Eq.\ \rf{orderzero} suggests
it must behave roughly as a functional linear 
in $E^{-2}_{nn} =n$,
and hence linear in the mass level.

It may also be possible to obtain other solutions of interest
by modifying the construction of $\cN$ in a different way.
In section V,
attention was primarily restricted to states of even excitation number.
However,
more general possibilities exist. 
For a bosonic tensor
$f_{mn}^{\la_1\ldots\la_m \mu_1\ldots\mu_n}$, 
we can separate the indices according to left and right sectors
so that $f_{mn}$ 
is annihilated by contracting with 
$(P_1)^{\pt{\mu_i}\mu_i'}_{\mu_i}$ and $(P_2)^{\pt{\la_i}\la_i'}_{\la_i}$.
The action of $L_0$ takes $f_{mn}$ to a combination
of terms having indices of the type
$(m-2,n)$, $(m-1,n-1)$, $(m,n-2)$,
$(m,n)$, $(m+1,n-1)$, $(m-1,n+1)$,
$(m+2,n)$, $(m+1,n+1)$, $(m,n+2)$.
These can be regarded as a field of black squares on a chessboard:
$(n+m)$ modulo 2 is conserved.
Since the star product combines 
an $(m,k)$ tensor with a $(k,n)$ tensor to 
yield an $(m,n)$ tensor,
it also preserves the sum of the indices modulo 2.
The possibility therefore exists of obtaining solutions 
$f_{mn}$ to the equations of motion restricted to even $(m+n)$
other than the nonperturbative vacuum $\cN$.

Yet another possibility is to consider solutions
mixing the colors on the chessboard.
Unlike solutions restricted to even $(m+n)$,
which in general may or may not violate Lorentz symmetry,
solutions with odd $(m+n)$ would have an odd number of oscillators
and hence would necessarily violate Lorentz symmetry.
In string field theory,
the occurrence of static interaction terms
quadratic in tensors and linear in scalars
suggests the possibility of spontaneous breaking of Lorentz symmetry
\cite{ks},
which would be accompanied by CPT violation 
when fields with odd level number are involved
\cite{kp}.
Indeed,
a family of Lorentz-violating and CPT-preserving
solutions in the open bosonic string
has been uncovered using the level-truncation scheme 
in the open bosonic string
\cite{kp2}.
It would be interesting to construct analytically the
states corresponding to these solutions. 

In general,
the topic of Lorentz-violating solutions appears
to be a potentially revealing subject for future study.
Relatively little is known about the nature of these solutions,
including whether they are stable.
Since they are spacetime independent,
it appears unlikely that they could correspond
to any of the lump solutions discussed in the literature to date.
The ubiquitous nature 
of the cubic scalar-tensor-tensor couplings
suggets that Lorentz-violating solutions are generic 
in string field theories with tachyons.
In fact,
the existence of scalar-tensor-tensor couplings
is not restricted to string field theory,
since they also arise in off-shell calculations of the action 
using renormalization-group methods in the string sigma model
\cite{kpp}.

Since the physical world is believed to include a tachyon 
in its spectrum at the electroweak scale (the Higgs field),
it is necessary for a realistic string theory 
to have at least one tachyonic mode.
If Lorentz-violating solutions are indeed generic in theories with tachyons,
it is natural to speculate that 
the four-dimensional Lorentz and CPT symmetries 
might also be spontaneously broken.
The resulting low-energy effects can be systematically studied
\cite{lowen}.
The exquisite sensitivity of modern experiments
to such effects 
\cite{kexpt,bexpt,lh,gg,rm,hd,db,rw,bh}
means that even effects suppressed
by the Planck scale could be detected.

Another interesting class of objects 
in string field theory is the set of lump solutions.
Using the level-truncation scheme,
it has been shown that bosonic D$p$-branes can be interpreted as unstable 
lump configurations of open-string tachyons on a D25-brane \cite{hk},
at least for large $p$.
Worldsheet-boundary renormalization-group methods
can be used to show 
that the mass of a tachyonic lump on a D$p$-brane
corresponds to that of a D$(p-1)$-brane \cite{hkm}.
This has also been verified numerically in a modified
level-truncation scheme incorporating derivative couplings 
\cite{msz}.
Some of these solutions are related to 
results in noncommutative field theory \cite{ew2}.
Various other calculations support these ideas
\cite{ms,nm}.

It is plausible that a variant of our methodology 
could be used to construct the lump solutions analytically.
In particular,
the construction of $\cN$ depends only 
on the formal structure of the vertex functions 
in terms of the matrices $C$ and $U$ 
with properties \rf{cuprop}.
These matrices emerge from 
the conversion between the $a_0$ oscillators
and the momentum representation in section IIA.
They originate in a set of matrices $C$ and $U'$
with identical formal properties \rf{cupprop}.
It therefore follows that any spacetime-independent solution
obtained in the momentum representation
has an inequivalent partner solution in the $a_0$ representation.
Moreover, 
the conversion from oscillator to momentum basis
can be performed in any number $p$ of spacetime dimensions desired.
The construction implemented for $\cN$ 
therefore suggests the existence of
string fields that are twisted squeezed states in
$(26-p)$ dimensions
and are independent of the remaining $p$ dimensions. 
We conjecture that these solutions
include the D$p$-branes,
along with instanton-type solutions.

Another topic of interest is the role and fate of the closed-string
modes in $\cN$.
The Zwiebach string field theory of the closed bosonic string
is nonpolynomial
\cite{bz}.
Consistent implementation of the level-truncation scheme 
therefore presents some difficulties.
At the level of the cubic interaction,
a nonperturbative solution appears to exist
in which the closed-string tachyon and the graviton
have no physical poles
\cite{kscbs}.
We anticipate that
a closed-string version of the squeezed field $\cS$
can be constructed for this solution.
If so,
its idempotency may make possible analytical study 
of the nonpolynomial action
in the squeezed basis.
It has also been shown that macroscopic closed strings can be regarded 
as solitons in the nonperturbative vacuum of the open bosonic string
\cite{hklm}.
It would be interesting to study solutions of this type
in the squeezed-state basis.

The methodology in this paper could also be applied to
study the nonperturbative vacua of 
other string field theories with tachyons,
including those for the non-GSO-projected superstring \cite{nb}
and the $p$-adic string
\cite{bfow}.
For the latter,
it has explicitly been proved that 
the extremum of the tachyon potential with a D25-brane 
is the vacuum without one
and that lump solutions are D-branes of lower dimensions
\cite{gs}.
In light of the possibility of obtaining exact results,
it would be interesting to investigate 
our construction for $\cN$ in this context.

\section*{Acknowledgments}
This work is supported in part 
by the United States Department of Energy
under grant number DE-FG02-91ER40661
and by the 
Portuguese Funda\c c\~ ao para a Ci\^ encia e a Tecnologia.

\begin{appendix}
\section{A useful identity}

This Appendix explicitly proves the identity \rf{bosonicr}
for multidimensional bosonic oscillators
under the conditions \rf{conditions}.
A similar method can be adopted to prove 
the result \rf{ghostriccati} 
for multidimensional ghost oscillators
subject to the constraints \rf{ghostconditions}.

The strategy we adopt is to seek 
a scalar function $\et(t)$ of a parameter $t$
and matrix functions 
$\al_{mn}(t)$, $\be_{mn}(t)$, $\ga_{mn}(t)$ 
of the same parameter $t$ such that the equation
\bea
&&\exp\left[t(a_m^\da A_{mn}a_n^\da+a_m^\da C_{mn}a_n+a_m B_{mn}a_n)\right]
\nonumber\\ &&~~
=\exp(\et)\exp(a_m^\da \al_{mn}a_n^\da)
\nonumber\\ &&\qquad\qquad
\times\exp(a_m^\da\ga_{mn}a_n)
\exp(a_m\be_{mn}a_n)
\eea
holds.
Here, the matrices $A$, $B$, and $C$ are assumed 
independent of $t$.
The idea is to take the derivative with respect to $t$ on both
sides of the equation, 
commute to the left all factors that appear,
and equate the corresponding terms on both sides.
This yields differential equations 
for the scalar and matrix functions of $t$.
Solving these yields the desired identity by setting $t=1$.

Adopting this procedure while keeping careful track of the index positions
yields the following set of matrix differential equations:
\bea
A&=&\dot\al-2\dot\ga\al+4\al \exp(-\ga^T)\dot\be \exp(-\ga)\al ,
\nonumber\\
B&=&\exp(-\ga^T)\dot\be \exp(-\ga) ,
\nonumber\\
C&=&\dot\ga-4\al\exp(-\ga^T)\dot\be \exp(-\ga) ,
\nonumber\\
0&=&\dot\et-2\,\hbox{Tr}[\exp(-\ga^T)\dot\be \exp(-\ga)\al ] ,
\label{eqs}
\eea
where $T$ denotes the transpose.
For simplicity in the above and in what follows,
we assume 
\bea
\al^T=\al,&&\qquad\al C^T=C\al,
\nonumber\\
\be^T=\be,&&\qquad\be C=C^T\be
\label{ansatz}
\eea
as a partial ansatz
that is to be verified after the solution is constructed.
The equations \rf{eqs} can be rearranged to give
\bea
\dot\al &=& A+2C\al+4\al B\al ,
\label{A}
\\
\dot\be &=& e^{\ga^T} B e^{\ga} ,
\label{B}
\\
\dot\ga &=& C + 4\al B , 
\label{C}
\\
\dot\et&=&2\,\hbox{Tr}(B\al) .
\label{et}
\eea
As initial conditions for these equations,
we take 
$\al(0) = \be(0) = \ga(0) = \et(0) = 0$.

Taking advantage of the given conditions \rf{conditions},
Eq.\ \rf{A} becomes 
\beq
\dot\al=(\al+\quarter CB^{-1})4B(\al+\quarter CB^{-1}) .
\eeq
Its solution is $\al+\quarter CB^{-1}=(-4Bt+Y)^{-1}$, 
where the initial condition fixes $Y=4BC^{-1}$.
This gives
\beq
\al(t)=
(1 - Ct)^{-1} At.
\eeq
Substituting this into Eq.\ \rf{C} gives 
$\dot\ga=-(t-C^{-1})^{-1}$,
which has solution 
\beq
\ga(t)=-\ln(1-Ct) .
\eeq
This in turn can be substituted into Eq.\ \rf{B},
leading to $\dot\be=B(1-Ct)^{-2}$.
The solution is
\beq
\be(t)=
Bt (1 - Ct)^{-1}.
\eeq
This leaves Eq.\ \rf{et}, which becomes 
\beq
\dot\et=\half\,\hbox{Tr}\left[\left((1-Ct)^{-1}-1\right)C\right]
\eeq
with solution
\beq
\et(t)=\half\,\hbox{Tr}\left[-\ln(1-Ct)-Ct\right] .
\eeq

We can now verify the ansatz \rf{ansatz}.
Finally,
setting $t=1$ yields the desired identity \rf{bosonicr}.

\end{appendix}

\end{multicols}

\begin{references}
\bibitem{fn2}
The application of these equations in later sections
is in the context of the Feynman-Siegel gauge
\cite{ws}.
This means that the usual zero modes of the ghost algebra 
can be disregarded here.
\bibitem{fn3}
The vacua $\ket{\pm}$ are defined following Ref.\ \cite{gj}.
For a single oscillator,
$\ket{\pm}$ are the ghost-number $\pm\half$ vacua
satisfying $\ket{+} = c_0 \ket{-}$, $\ket{-} = b_0\ket{+}$.
\end{references}
\end{document}